\documentclass[12pt,preprint]{aastex}

\input psfig.tex

\begin{document}

\title{A parameter-free statistical measurement of halos with power
   spectra}

\author{Ping He\altaffilmark{1,2}, Long-Long Feng\altaffilmark{1,3}
 and Li-Zhi Fang\altaffilmark{2}}

\altaffiltext{1}{National Astronomical Observatories, Chinese
Academy of Science, Chao-Yang District, Beijing, 100012, P.R.
China}

\altaffiltext{2}{Department of Physics, University of Arizona,
Tucson, AZ 85721}

\altaffiltext{3}{Purple Mountain Observatory, Nanjing 210008,
P.R. China}

\begin{abstract}

We show that in the halo model of large-scale structure formation,
the difference between the Fourier and the discrete wavelet
transform (DWT) power spectra provides a statistical measurement
of the halos. This statistical quantity is free from parameters
related to the shape of the mass profile and the identification
scheme of the halos. That is, the statistical measurement is
invariant in the sense that models with reasonably defined and
selected parameters of the halo models should yield the same
difference of the Fourier and DWT spectra. This feature is useful
to extract ensemble-averaged properties of halos, which cannot be
obtained with the identification of individual halos. To
demonstrate this point, we show with WIGEON hydrodynamic
simulation samples that the spectrum difference provides a
quantitative measurement of the discrepancy of the distribution of
baryonic gas from that of the underlying dark matter field within
halos. We also show that the mass density profile of halos in
physical space can be reconstructed with this statistical
measurement. This profile essentially is the average over an
ensemble of halos, including well-virialized halos, as well as
halos with significant internal substructures. Moreover, this
reconstruction is sensitive to the tail of the mass density
profile. We showed that the profile with a $1/r^3$ tail gives very
different result from that with $1/r^2$. Other possible
applications of this method are discussed as well.

\end{abstract}

\section{Introduction}

The dynamical equations of cosmic collisionless particles (dark
matter) do not have preferred scales. Therefore, if the initial
density perturbations are Gaussian and scale-free, the equations
admit a self-similar solution, and the cosmic clustering of dark
matter has to be scale-invariant (Peebles 1980). To sketch this
clustering, the halo model assumes that the cosmic mass field in
the nonlinear regime is given by a superposition of the halos on
various spatial scales. The halo-halo correlation function on
scales larger than the halo size is given by the two-point
correlation function of the initially linear Gaussian field. In
the early version of the halo model, the mass density profile of
the halos is supposed to be self-similar. For instance, in the
model of Neyman \& Scott (1952), the distribution of galaxies
within various clusters is assumed to obey the same probability
law, which is independent of the mass of clusters. Scherrer \&
Bertschinger (1991) used a scale-free function $g(x)$ to model the
density profile $f(m,x)=mg(x)$ of a mass $m$ halo. In the current
halo model, the halos are generally described by the universal
mass profiles. There are several different versions of the
profiles (Hernquist 1990; Navarro et al. 1996; Moore et al. 1999).
A common feature of these halo density profiles is that the
parameters used in the density profiles are either
mass-independent or dependent on mass approximately by a power
law. Therefore these density profiles essentially scale with
respect to halo mass.

The halo model has been extensively applied to modeling the
hierarchical formation of galaxies and other objects (e.g., Wang
et al. 2004 and references therein). It tries to explain the
formation and evolution of galaxies with the mass function, the
universal density profile, two-point correlations of host halos,
and the bias model of the relevant objects (e.g., Cooray \& Sheth
2002 and references therein).

The basic assumptions of the halo model are correct only
statistically. Previous $N$-body simulations show that the shapes
of halos are often nonspherical and that the parameters of mass
profiles of halos are dependent on the halo mass. For instance,
probably no more than 70\% of halos can be fitted by the universal
mass profile. There is a considerable amount of variation in the
density profile even among the halos that can be fitted by the
universal mass profile (Jing 2000; Bullock et al. 2001). The
universality of spherically symmetric profiles is approximately
valid for the average over ensembles of halos but not so for
individual halos. Moreover, the mass function of halos also
depends on halo finders (Jenkins et al. 2001). That is, the
ensembles of halos with a given mass do not completely depend on
mass but also depend on the parameters used in the halo
identification. This may lead to assigning different properties to
halos at a given mass because the masses are assigned using a
different algorithm. Thus, it would not be trivial in comparing
halo catalogs constructed using two different halo identification
schemes.

Obviously, the parameters of the halo finder should be irrelevant
to physical conclusions yielded from the halo model. Therefore, it
would be significant to directly measure the ensemble-averaged
properties of halos without invoking {\em a priori} parameters
related to the details of mass profiles and the specific halo
finders. In this paper, we propose a parameter-free statistical
scheme to measure the halos. It relies only on the physical
essence of the halo model and not on the assumptions of the shape
of the halos. This measurement is based on a relation between the
power spectra of Fourier and discrete wavelet transform (DWT)
bases. For Gaussian fields with either colored or white spectra,
the Fourier and the DWT power spectra are equivalent to each
other; that is, the Fourier power spectrum $P(n)$ can be
transferred to the DWT power spectrum $P_j$ and vise versa. Yet,
for non-Gaussian fields, the Fourier and the DWT power spectra are
no longer equivalent. The difference between the Fourier and the
DWT power spectra completely comes from the non-Gaussianity of the
field. Thus, if the non-Gaussianity of a mass field is caused by
halos, the difference between the Fourier and the DWT power
spectra provides a direct statistical measurement of halos.

The paper is organized as follows. In \S 2, we derive the relation
between the Fourier and the DWT power spectra and show that the
difference between the Fourier and the DWT power spectra provides
the ensemble-averaged measurement of halo profiles. \S 3 describes
the samples used for demonstration. \S 4 demonstrates the possible
applications of this measure, including (1) the distribution of
baryonic gas in halos and (2) the reconstruction of the halo mass
profile. The discussion and conclusion are given in \S 5.

\section{Theory}

\subsection{The Fourier and DWT power spectra}

To show the essence of the method, we consider a one-dimensional
random density distribution $\rho(x)$ in the range $0<x<L$. It is
straightforward to generalize the result to two- or
three-dimensional fields. The Fourier power spectrum is given by
\begin{equation}
P(n)= \langle |\delta_n|^2\rangle,
\end{equation}
where $\delta_n = (1/L)\int_0^{L}\delta(x)\exp(-i2\pi nx/L)dx$ and
$\langle ...\rangle$ stands for ensemble average. Similarly, the
DWT power spectrum is given by
\begin{equation}
P_j= \langle|\tilde{\epsilon}_{j,l}|^2\rangle,
\end{equation}
where $\tilde{\epsilon}_{j,l}=\int_{0}^{L} \delta(x)
\psi_{j,l}(x)dx$, where $\psi_{j,l}(x)$ is the wavelet function
with scale index $j$ and position index $l$ (All the wavelet
notations used in this paper are the same as in Fang and Feng
[2000]).  For a statistically homogeneous Gaussian field, the
right hand side of equation (2) is $l$-independent.

The Fourier variable $\delta_n$ can be expressed by the DWT
variable $\tilde{\epsilon}_{j,l}$ as (Fang \& Feng 2000)
\begin{equation}
\delta_n  =  \frac{1}{L}\sum_{j=0}^{\infty} \sum_{l=0}^{2^j-1}
\tilde{\epsilon}_{j,l} \hat{\psi}_{j,l}(n) \\ \nonumber
  =  \sum_{j=0}^{\infty}  \sum_{l=0}^{2^j-1}
\left (\frac{1}{2^{j}L}\right )^{1/2} \tilde{\epsilon}_{j,l}
e^{-i2\pi nl/2^j} \hat{\psi}(n/2^j), \ \ \ \ \ \ \ \ \ n \neq 0.
\end{equation}
where $\hat{\psi}$ is the Fourier transform of the basic wavelet
function. Therefore, the Fourier power spectrum is given by
\begin{equation}
P(n) =
\frac{1}{L^2}
\sum_{j=0}^{+\infty}\sum_{l=0}^{2^j-1}
\sum_{j'=0}^{+\infty}\sum_{l'=0}^{2^{j'}-1}
\langle\tilde{\epsilon}_{j,l}\tilde{\epsilon}_{j',l'}\rangle
\hat{\psi}_{j,l}(n)\hat{\psi}^{\dagger}_{j',l'}(n)
\end{equation}
For Gaussian fields,
$\langle\tilde{\epsilon}_{j,l}\tilde{\epsilon}_{j',l'}\rangle=
P_j\delta_{jj'}\delta_{ll'}$, we have
\begin{equation}
P(n) =\frac{1}{L} \sum_{j=0}^{\infty}P_j
 \left |\hat{\psi} \left (\frac {n}{2^j}\right) \right |^2.
\end{equation}
Therefore, the Fourier power spectrum, $P(n)$, and the DWT power
spectrum, $P_j$, are equivalent for Gaussian fields.

Generally, equations (1)--(3) yield
\begin{eqnarray}
\lefteqn{ P(n) - \frac{1}{L} \sum_{j=0}^{\infty}P_j
 \left |\hat{\psi} \left (\frac {n}{2^j}\right) \right |^2 = }\\
   \nonumber
& &  \sum_{j=0}^{+\infty}\sum_{l, l'=0, l\neq l'}^{2^j-1}
 \langle\tilde{\epsilon}_{j,l}\tilde{\epsilon}_{j,l'}\rangle
  \hat{\psi}_{j,l}(n)\hat{\psi}^{\dagger}_{j,l'}(n)  +
  \sum_{j, j'=0, j'\neq j}^{+\infty}
  \sum_{l=0}^{2^j-1}\sum_{l'=0}^{2^{j'}-1}
\langle\tilde{\epsilon}_{j,l}\tilde{\epsilon}_{j',l'}\rangle
  \hat{\psi}_{j,l}(n)\hat{\psi}^{\dagger}_{j',l'}(n)
\end{eqnarray}
The two terms on the right-hand side originate from the
non-Gaussianity of the field. That is, the difference between the
Fourier and DWT power spectra is a measure of the non-Gaussianity
of the field.

\subsection{The Fourier and DWT power spectra in the halo model}

In the halo model, the developed mass density field $\rho(x)$ is
assumed to be a superposition of halos on various scales. It is
\begin{equation}
\rho(x) = \sum_i \rho_i(x - x_i) = \sum_i m_i u( x - x_i, m_i)
= \int d x' \sum_i m_i\delta^D(x'-x_i)u(x - x', m_i),
\end{equation}
where $m_i$ is the mass of halo $i$, $\delta^D(x)$ is the Dirac
delta function, and $u(x - x_i, m_i)$ is the normalized mass
density profile of halo $i$, i.e., $\int dx \rho_i(x - x_i)=\int
dx m_i u(x - x_i, m_i)=m_i$. In this case, the DWT variable is
\begin{equation}
\tilde{\epsilon}_{j,l} = (1/\bar{\rho})\sum_i m_i \int dx u(x -
x_i, m_i)\psi_{ j,l}(x).
\end{equation}
The covariance
$\langle\tilde{\epsilon}_{j,l}\tilde{\epsilon}_{j,l'}\rangle$ is
then given by
\begin{equation}
\langle \tilde{\epsilon}_{j,l}\tilde{\epsilon}_{j,l'}\rangle
=\langle \tilde{\epsilon}_{j,l}\tilde{\epsilon}_{j,l'}\rangle^h
  +
\langle \tilde{\epsilon}_{j,l}\tilde{\epsilon}_{j,l'}\rangle^{hh},
\end{equation}
where the first and second terms on the right-hand side are
usually called one and two halo terms, respectively (Cooray \&
Sheth 2002). The two-halo term is given by the two-point
correlation function of the Gaussian perturbations, and therefore,
$\langle \tilde{\epsilon}_{ j,l}\tilde{\epsilon}_{j',l'}
   \rangle^{hh}\simeq 0$, when $l\neq l'$, and the one-halo term,
$\langle
\tilde{\epsilon}_{j,l}\tilde{\epsilon}_{j,l'}\rangle^h$, is (Feng
\& Fang 2004)
\begin{equation}
\langle \tilde{\epsilon}_{j,l}\tilde{\epsilon}_{j,l'}\rangle^h
   =
\int  dm n(m)\left ( \frac{m}{\bar{\rho}} \right )^2\int dx_1 u({
x}_1, m) \int d{x} \int d{x'}\psi_{j,l}({x})u({ x_1+ x- x'}, m)
\psi_{j,l'}({x'}),
\end{equation}
where $n(m)$ is the number density of halos with mass $m$. Thus,
we have
\begin{eqnarray}
\langle \tilde{\epsilon}_{j,l}\tilde{\epsilon}_{j,l'}\rangle &  = &
\langle \tilde{\epsilon}_{ j,l}\tilde{\epsilon}_{j,l'}\rangle^h \\
   \nonumber
  &  = &
 \frac{1}{2^j}\sum_{{n}=-\infty}^{\infty}
   |\hat{\psi}(n/2^j)|^2\cos[2\pi n(l-l')/2^j]
  \int dm n(m)\left ( \frac{m}{\bar{\rho}} \right )^2|\hat{u}(n, m)|^2
\end{eqnarray}
where $\hat{u}({n}, m)=\int_0^L u(x, m)\exp(-i 2\pi n x/L)dx$ is
the Fourier transform of the normalized mass profile.

The second term on the right-hand side of equation (6) contains a
factor $\hat{\psi}_{j,l}(n)\hat{\psi}^{\dagger}_{j',l'}(n) \propto
\hat{\psi}(n/2^j)\hat{\psi}^{\dagger}(n/2^{j'}) \simeq 0$, because
$n/2^{j}$ and $n/2^{j'}$ differ by a factor of $2^{|j-j'|}$ if
$j\neq j'$, and then, the non-zero ranges of $\hat{\psi}(n/2^j)$
and $\hat{\psi}^{\dagger}(n/2^{j'})$ do not overlap. Thus,
equation (6) yields
\begin{equation}
P(n) -\frac{1}{L} \sum_{j=0}^{\infty}P_j
 \left |\hat{\psi} \left (\frac {n}{2^j}\right) \right |^2
  =L \left (\int dm n(m)\left ( \frac{m}{\bar{\rho}} \right )^2|
  \hat{u}({n}, m)|^2 \right ) \sum_{j=0}^{\infty}
   \left |\hat{\psi}\left (\frac{n}{2^j} \right ) \right |^4,
\end{equation}
where we have used
$\sum_{l=0}^{2^j-1}e^{-i2\pi(n-n')l/2^j}=2^j\delta_{n,n'}$.
Equation (12) is the major theoretical result of this paper. Since
we know $\sum_{j=0}^{\infty}|\hat{\psi}(n/2^j)|^4$, from the
difference between the Fourier and DWT power spectra, one has as a
statistical measure
\begin{equation}
U^2(n)= \int dm n(m)\left ( \frac{m}{\bar{\rho}}
\right)^2|\hat{u}({n}, m)|^2.
\end{equation}
Since $n(m)$ is the number density of $m$ halos, $U^2(n)$ is an
average of the Fourier mode $n$ of mass profile $|\hat{u}({n}, m)|^2$
over the ensemble of halos, and the weight of the average is
$(m/\bar{\rho})^2$.

\subsection{Constraint on halo models with power spectrum}

Note that the left hand side of eq.(12) depends only on power
spectra of the Fourier modes $P(n)$ and DWT modes $P_j$,
regardless of the definition and selection of particular
parameters used in the halo profile and halo mass function. On the
other hand, the right hand side of eq.(12) depends on the mass
function $n(m)$ and mass profile $\hat{u}({n}, m)$ of halos.
Therefore, $U^2(n)$ should be invariant with respect to the
parameters used for the definition and selection of halos. Thus,
$U^2(n)$ provides a general constraint on models of the mass
function and mass profile of halos. A reasonable model of the mass
function and mass profile of halos must satisfy relation (12). The
statistic $U^2(n)$ can be used to compare models with different
definitions and selections of halos.

For instance, the universal mass profile of halos generally is
assumed to be
\begin{equation}
\rho(r,m)=\frac{\rho_s}{(r/r_s)^{\alpha}(1+ r/r_s)^{\beta}}.
\end{equation}
where the parameters $r_s$ and $\rho_s$ are, respectively, the
scale radius and characteristic density of the halo. The index
$(\alpha,\beta)$ is mass independent. It can be $(1,3)$ of
Hernquist (1990), $(1,2)$ of Navarro et al. (1996), or $(3/2,3/2)$
of Moore et al. (1999). Obviously, models with different indices
$(\alpha,\beta)$ yield different $n$-dependences of $\hat{u}({n},
m)$. Thus, if one uses the same mass function $n(m)$ for all
models $(\alpha,\beta)$, $U^2(n)$ will be $(\alpha,\beta)$
dependent. This is unacceptable. Therefore, in order to match the
constraint in eq.(12), for different indices $(\alpha,\beta)$, we
must use different mass function $n(m)$. In other words, in the
halo model, the mass function $n(m)$ of halos actually is
halo-shape-parameter dependent. Similarly, the mass function
$n(m)$ also depends on the algorithms of halo identification. This
point is well-understood in the halo model (Jenkins et al 2001).
However, eq.(12) provides an effective tool to test whether the
parameters of the halo and halo finder yield a consistent set of
halo model ingredients. In other words, the invariance of $U^2(n)$
with respect to the parameters sets a necessary condition of the
consistent set of the definition and selection of halo parameters.

\section{Samples}

To demonstrate the possible application of the statistic $U(n)$,
we use the cosmological hydrodynamic simulation samples used in He
et al. (2004) and Pando et al. (2004). The samples are produced by
the WIGEON (Weno for Intergalactic medium and Galaxy Evolution and
formatiON) code, which is a cosmological hydrodynamic/$N$-body
code based on the Weighted Essentially Non-Oscillatory (WENO)
algorithm (Harten et al. 1986; Liu, Osher, \& Chan 1994; Jiang \&
Shu 1996; Shu 1998; Fedkiw, Sapiro \& Shu 2003; Shu 2003). The
WENO schemes use the idea of adaptive stencils in the
reconstruction procedure based on the local smoothness of the
numerical solution to automatically achieve high order accuracy
and non-oscillatory properties near discontinuities. Therefore,
the WENO algorithm provides a high order precision simulation for
fluid with strong shocks.

In a system in which the baryonic gas is gravitationally coupled
with dark matter, the temperature of the baryonic gas is generally
in the range $10^{4-6}$ K, and the speed of sound in the baryonic
gas is only a few km s$^{-1}$ to a few tens km s$^{-1}$. On the
other hand, the rms bulk velocity of the baryonic gas is on the
order of hundreds km s$^{-1}$ (Zhan \& Fang 2002). Hence, the Mach
number of the gas can be as high as $\sim$100 (Ryu et al. 2003).
The shocks and discontinuities are extremely strong. The WENO code
for cosmology has passed  necessary tests including the Sedov
blast wave and the formation of the Zeldovich pancake (Feng, Shu,
\& Zhang 2004). It has been successfully used to produce the QSO
Ly$\alpha$ transmitted flux (Feng, Pando, \& Fang 2003). The
statistical features of these samples are in good agreement with
observed features not only on second order measures, like the
power spectrum, but also up to orders as high as eighth for the
non-Gaussian behavior (Pando, Feng, \& Fang 2002). Hence we
believe that the WENO samples would be suitable to study the
non-Gaussian feature of the mass field with the statistic $U(n)$.

The simulations were performed in a periodic, cubic box of size 25
$h^{-1}$Mpc with a 192$^3$ grid and an equal number of dark matter
particles. The simulations start at a redshift $z=49$. The linear
power spectrum is taken from the fitting formulae presented by
Eisenstein \& Hu (1998). The baryon fraction
$f_c=\bar{\rho}_b/\bar{\rho}_{dm}=\Omega_{b}/\Omega_{dm}$ is fixed
using the constraint from primordial nucleosynthesis as
$\Omega_b=0.0125h^{-2}$ (Walker et al. 1991). A uniform
UV-background of ionizing photons is assumed to have a power-law
spectrum of the form $J(\nu) =J_{21}\times10^{-21}
(\nu/\nu_{HI})^{-\alpha}$erg s$^{-1}$cm$^{-2}$sr$^{-1}$Hz$^{-1}$,
where the photoionizing flux is normalized by the parameter
$J_{21}$ at the Lyman limit frequency $\nu_{HI}$ and is suddenly
switched on at $z > 10$ to heat the gas and reionize the universe.
The atomic processes, including ionization, radiative cooling and
heating, are modeled in the same way as in Cen (1992) in a plasma
of hydrogen and helium of primordial composition ($X=0.76$,
$Y=0.24$). The processes such as star formation, and feedback due
to SN and AGN activities have not yet been taken into account.

For statistical studies, we randomly sampled 500 one-dimensional
fields from the simulation results at redshift $z=0$. Each
one-dimensional sample, of size $L$=25 $h^{-1}$Mpc, contains 192
data points, every one of which contains mass densities of dark
matter and baryonic gas.

\section{Applications}

\subsection{The discrepancy of baryonic gas from the dark matter}

It has been well known that in the nonlinear regime the clustering
of baryonic gas will decouple from the underlying dark matter
(e.g., He et al. 2005; Kim et al 2005). For instance, Figure 1
plots two-dimensional contours of the baryonic gas and dark matter
densities. One can see a halo of the scale $\sim$5 $h^{-1}$ Mpc at
the lower part of the plots. The size of the
$\rho_{b}/\bar{\rho_b}>1$ region is obviously larger than the dark
matter counterpart at $\rho/\bar{\rho}>1$. It means that much
baryonic gas remains outside the halo. This is a discrepancy of
baryonic gas from the dark matter. Although the discrepancy can be
directly seen by eye, a quantified measure is still necessary. The
discrepancy is significant in massive halos. One can have a
quantified measure of the discrepancy of the massive halos.
However, a measure of the discrepancy over the whole field is also
necessary. The latter can be done with the statistic in eq.(12).

Similar to eq.(7), the mass field of baryonic gas in the halo model
can be written as
\begin{eqnarray}
\rho_b(x)& = &\sum_i \rho_{i,b}(x - x_i) =
  \sum_i m_if(m_i) u^b( x - x_i, m_i)
    \\ \nonumber
 & = & \int d x' \sum_i m_i f(m_i)
   \delta^D(x'-x_i) u_b(x - x', m_i),
\end{eqnarray}
where $\rho_{i,b}(x - x_i)$ is the mass profile of baryonic gas in
the halo $i$, $u_b( x - x_i, m_i)$ is the normalized baryon mass
profile of the halo with dark matter mass $m_i$, i.e. $\int dx
u_b(x - x_i, m_i)=1$, and $f(m_i)=m_b/m_i$ is the baryon fraction
of the halo with mass $m_i$. Thus, with a similar procedure as in
\S 2, we can show that the difference between the Fourier and DWT
power spectra of the mass field in eq.(15) gives a measurement of
baryonic gas in halos as
\begin{eqnarray}
U^2_b(n) & = & \int dm n(m)\left (\frac{m_b}{\rho_{b}} \right
)^2|\hat{u}_b({n}, m)|^2\\
\nonumber & & \int dm n(m) \left ( \frac{m}{\rho} \right)^2\left
(\frac{f(m)}{f_c}\right )^2|\hat{u}_b({n}, m)|^2,
\end{eqnarray}
where $\hat{u}_b({n}, m)$ is the Fourier transform of $u_b(x)$,
and $f_c=\Omega_{b}/\Omega_{dm}$.

If the distribution of baryonic gas is given by a similarity
mapping of dark matter (e.g., Kaiser 1986), i.e.
$\rho_{b}(x)=f_c\rho(x)$, we will have $U^2(n)=U^2_b(n)$. Thus,
any difference between $U^2(n)$ and $U^2_b(n)$ will indicate the
violation of the similarity mapping between baryon gas and dark
matter. The difference may originate from the two factors: (1) the
normalized mass profiles $u_b(x)$ are different from those of the
dark matter $u(x)$, and (2) the baryon fraction $f(m)$ is not
always equal to the cosmic value $f_c$, but is $m$-dependent.

For the hydrodynamic simulation samples of \S 3, the results of
$U_{b}(n)$ and $U(n)$ are shown in Figure 2. The error bars in
Figure 2 indicate the 68\% confidence level estimated by the
bootstrap method. Figure 2 shows that $U_{b}(n)$ is always lower
than $U(n)$ in the range $n\geq 2$. For each data point at $n$,
the mean values of $U_{b}(n)$ and $U(n)$ are also derived by
bootstrap re-sampling from the 500 samples. Thus, we can do a
$t$-test at every $n$ to assess the significance of the deviation
of $U(n)$ from $U_b(n)$. We find that the significance of
$U_{b}(n)$ to be different from $U(n)$ is greater than 99\% with a
$t$-test for every $n$ in the range 5 to 15. The result $U_{b}(n)
< U(n)$ means that the mean baryon fraction on scale $L/n$ should
be lower than $f_c$. Therefore, from Figure 2 one can conclude
that the baryon fraction within a few Mpc surrounding a halo is
less than $f_c$. Thus, more baryonic matter remains outside the
collapsed objects. This point can be seen clearly from the right
panel of Figure 3, which shows that $f/f_c$ is lower than 1 at the
central part of this halo, while the contours of $f/f_c>1$ are
outside the central part. Observations also showed that the baryon
fraction in galaxy clusters seems to be lower than the result of
primordial nucleosynthesis (Ettori et al. 2004).

As mentioned above, the statistic $U$-$U_b$ gives a quantitative
measurement of this feature with the entire mass field, not from
individual halos. Moreover, if the relation between profiles of
baryonic gas $\rho_{b}(x)$ and dark matter $\rho(x)$ can be
described by a nonlocal bias model, i.e. $\rho_{b}(x)=\int
b(x-x')\rho(x')dx$, where $b(x-x')$ is the bias function, we have
$U_b(n)=b(n)U(n)$. Therefore the statistic $U$-$U_b$ can be used
to detect the biased distribution of baryonic gas, and the
ensemble averaged nonlocal bias function in Fourier space.

\subsection{Reconstruction of mass density profiles}

We now consider the problem of recovering the mass profile. Since
$U(n)$ is weight-averaged, from $U(n)$ we cannot get
$\hat{u}(n,m)$ for each given mass. However, the weight factor is
$n(m)m^3d\ln m\propto
[m/m_{col}(z)]^{3-\mu}\exp[-m/m_{col}(z)]d\ln m$, and $3-\mu >1$.
The maximum of the weight is at $\simeq m_{col}(z)$. Hence, $U(n)$
is approximately equal to $|\hat{u}({n}, m)|$, where the logarithm
mass $\ln m$ is within the range $\ln m_{col} \pm \Delta \ln m$,
and $\Delta \ln m \simeq 1$.

We can reconstruct the mass density profile $u(x)$ by a Fourier
transform as
\begin{equation}
u(x)\propto U(x)=\frac{2}{L}\sum_{n=1}^{N}U(n)\cos(2\pi n x/L).
\end{equation}
In eq. (17) $N$ is dependent on the sample resolution. Since
$u(x)= 0$ at large $x$, and therefore, $U(n=0) =0$. Using the
$U(n)$ of dark mater shown in Figure 2, we have $U(x)$ given in
Figure 4. The error bar is also estimated by the bootstrap method.
First, we see that in Figure 4, there is a region with
non-physical density $U(x)<0$. This is because of the absence of
the power on scales that are comparable to or larger than the size
of the simulation box. Second, we do not have the information of
the mass profile on scales less than 0.2 $h^{-1}$ Mpc, which is
limited by the resolution of the sample. Nevertheless, Figure 4
already shows some interesting properties, which are different
from those of the mass density profile given by an average of
individual halos.

First, we fit $U(x)$ by the mass profile of Eq.(14). For
$(\alpha,\beta)= (1,2)$ and $(1,3)$, the corresponding one-dimensional
profile is then
\begin{equation}
u(x) \propto \frac{1}{(1+ |x/r_s|)^{\beta-\alpha}}.
\end{equation}
The fitting results are shown in Figure 4. The values of scale
radii $r_s$ are, obviously, in the sense of a statistical average
over the field. It should be pointed out that the fitting done in
Figure 4 is to cover a range as large as 6 h$^{-1}$ Mpc. This is
because the $U(x)$ provides a measurement of the entire mass
field. The fitting profile is also an ensemble averaged
measurement of the entire mass field.

Figure 4 shows that the fitting curves for $(\alpha,\beta)= (1,2)$
and $(1,3)$ are about the same. That is, both fitting curves have
about the same goodness-of-fit. However, the parameters of $r_s$
for $(\alpha,\beta)= (1,2)$ and $(1,3)$ are very different. The
former is 0.2 $h^{-1}$ Mpc, while the latter is 0.7 h$^{-1}$ Mpc.
For the model $(\alpha,\beta)=(1,2)$, the scale radii of halo with
mass $m_{col}\simeq 10^{13} M_{\sun}$, is found to be $r_s \simeq
0.06$ h$^{-1}$ Mpc (Bullock et al. 2001). It is much smaller than
0.2 $h^{-1}$ Mpc. This is probably because that the number 0.2
$h^{-1}$ Mpc is close to the sample resolution. We may miss some
halos with small $r_s$. However, there is also a physical reason
leading to $r_s > 0.06$ h$^{-1}$ Mpc. The profile of halos is
found to be sensitive to the dynamical status (Jing 2000). With an
unbiased selection of all massive halos from $N$-body simulations,
it has found that, for the LCDM model, there are only about 44\%
of the halos than can be fitted by the profile (14) of $(\alpha,
\beta)=(1,2)$ with a fitting residual dvi$_{max}<0.15$. These
halos are well virialized. About 40\% of halos have $0.15 <$
dvi$_{max}<0.30$, and 16\% have dvi$_{max}>0.30$. The latter has
obvious sub-structures. The parameter $r_s$ increases with
dvi$_{max}$. The value of $r_s$ of dvi$_{max}>0.3$ halos are
larger than those of dvi$_{max}<0.15$ halos by a factor of about
2. Therefore, in the case of $(\alpha,\beta)=(1,2)$, the parameter
of mass profile given by the $U(n)$ recovery is basically
reasonable. It is an ensemble averaged parameter over virialized
as well as sub-structured halos.

On the other hand, for the case of $(\alpha,\beta)=(1,3)$,
$r_s=0.7 $ h$^{-1}$ Mpc is too large. This is probably because
that the measure $U(n)$ is over the whole mass field, and
therefore, it is more sensitive to the tail of the profile
eq.(18). The tail of the $(\alpha,\beta)=(1,3)$ profile, $1/r^4$
(or in one dimension $1/x^2$) is more steep than that of
$(\alpha,\beta)=(1,2)$, $1/r^3$ (or in one-dimension $1/x$). Thus,
the fitting by $(\alpha,\beta)=(1,3)$ yields a large value of
$r_s$. Therefore, the $U(n)$ recovery of the mass profile
indicates that the profiles  $(\alpha,\beta)=(1,3)$ may not be
representative of LCDM halos.

\section{Discussions and conclusions}

In the Fourier representation, the essential difference between
the Gaussian field and the halo-filled field is in the
distribution of phases in Fourier modes. As coherent structures,
the phases of the density perturbations of halos are highly
correlated, while the phases of the Gaussian field are random, or
uncorrelated. On the other hand, the covariance of the Fourier
modes is not sensitive to the phases of perturbations, while the
DWT covariance contains information about both scale and position,
which is phase-sensitive. Therefore, even with second-order
statistics of Fourier and DWT modes (in the current case, it is
the difference between their power spectra), one can effectively
draw information about halos.

We show that, in the halo model of large-scale structure
formation, the difference between the Fourier and the DWT power
spectra provides a statistical measurement of the halos. This
method has all the advantages of power spectrum statistics, since
it directly reflects the physical scales of the clustering
processes that affect structure formation. Mathematically, the
positive definiteness of the power spectrum is useful for
constraining the parameter space. Moreover, it would also be easy
to estimate the effect of redshift distortion. For instance, the
linear redshift distortions of the Fourier and the DWT power
spectra are the same (Yang et al. 2001; Zhan \& Fang 2003).
Therefore, the Fourier - DWT power spectrum difference is free
from linear redshift distortions. In the halo model, non-linear
redshift distortions come from halos. Thus, the Fourier - DWT
power spectrum difference depends only on the redshift distortions
given by halos. This would be useful for studying the velocity
profile of halos.

This statistic is invariant with respect to various models with
reasonably defined and selected parameters of the halo models.
Therefore, it is a constraint for a consistent set of halo model
ingredients. This feature is useful for extracting information of
halos, containing both virialized halos and halos with significant
sub-structures. This cannot be done with the identification of
halos. This statistic measurement is on the whole map of various
structures, such as galaxies, quasars, Ly-alpha absorption
transmission, etc., and therefore, it would be easy to handle
various selections.

With the measurement of $U(n)$, we may investigate the bias of
baryonic gas with respect to dark matter. In particular, the two
types of the bias, (1) $u_b(x)\neq u(x)$, and (2) $f_c(m)\neq f_c$
would be distinguishable with $U(n)$. This method can also be used
to detect the difference of spatial distributions of different
objects of the universe.

In this paper, we only considered one-dimensional problems. In the
three-dimensional case, the DWT power spectrum contains both
diagonal and non-diagonal modes. It has been shown that the power
spectra of the diagonal and non-diagonal modes are effective in
detecting different clustering feature (Yang et al 2001).
Therefore, the difference between three-dimensional Fourier and
DWT power spectra would provide more constraints and information
on halo models of the cosmic mass field in the nonlinear regime.

\acknowledgments

We thank the anonymous referee for a stimulating report. P.H. is
supported by a Fellowship of the World Laboratory. L.L.F.
acknowledges support from the National Science Foundation of China
(NSFC) and the National Key Basic Research Science Foundation.
This work is partially supported by the National Natural Science
Foundation of China (10025313) and the National Key Basic Research
Science Foundation of China (NKBRSF G19990752).

\newpage

\begin{figure*}[hbt]
\centerline{\psfig{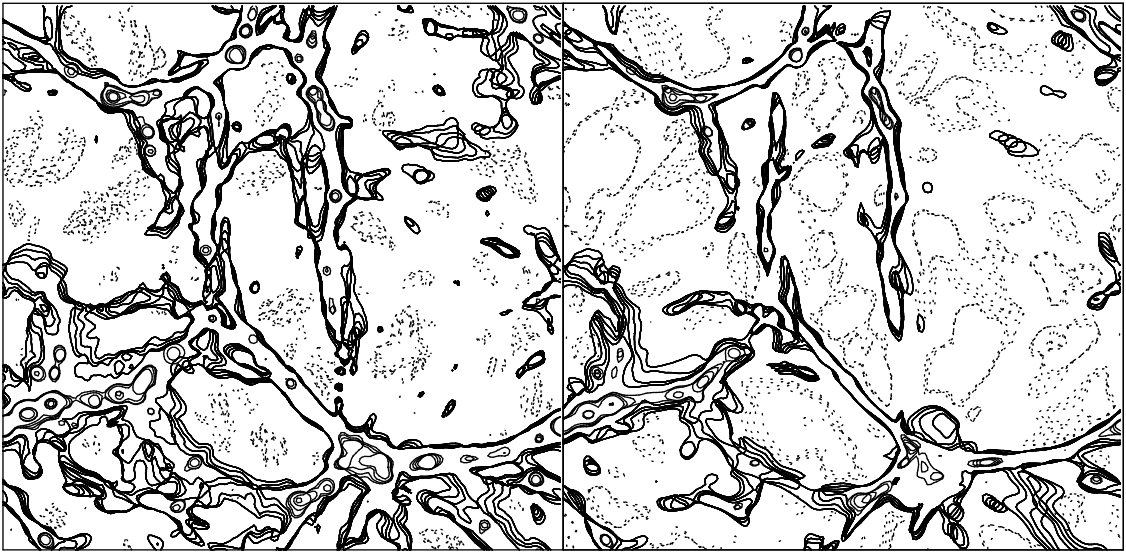}}

\caption{Baryonic gas (left) and dark matter (right)
         density contour plots for a slice of 0.26
         h$^{-1}$ Mpc thickness at $z=0$. The solid contours
         represent overdense regions with $\rho/\bar{\rho}>1$.
         The dotted lines represent underdense regions with
         $\rho/\bar{\rho}<1$ }
\end{figure*}

\newpage
\begin{figure*}[hbt]
\centerline{\psfig{figure=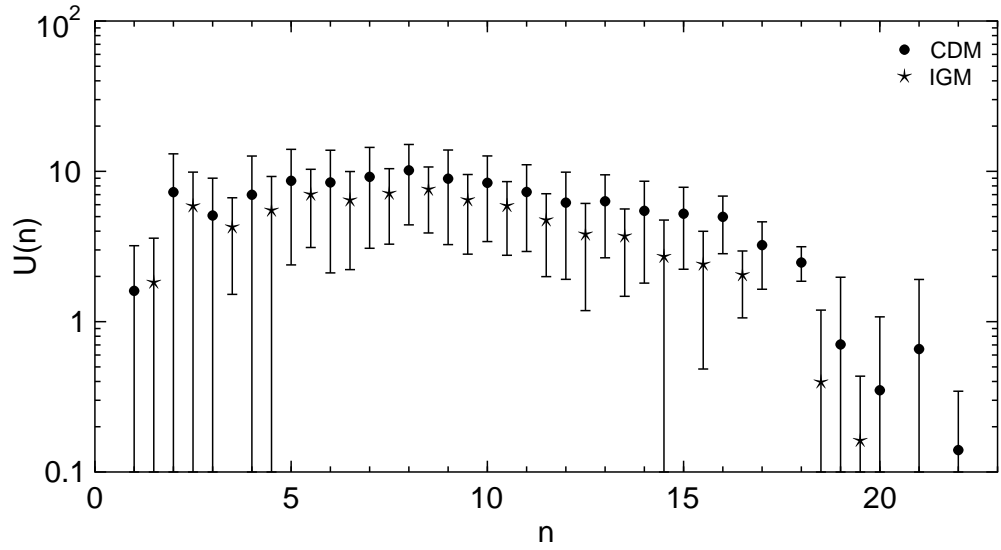}}

\caption{The statistical measurements $U(n)$ (filled
         circle) and $U_b(n)$ (star) versus n. For clarity,
         $U_b(n)$ is right-shifted for half unit. The error
         bars are estimated by the bootstrap method}
\end{figure*}

\newpage
\begin{figure*}[hbt]
\centerline{\psfig{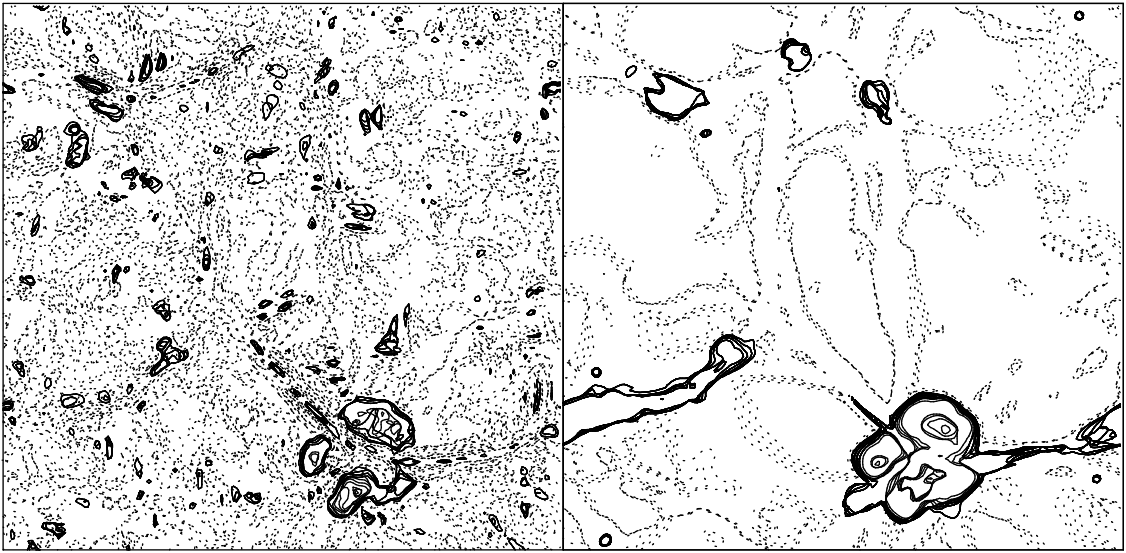}}

\caption{Temperature (left) and baryon-to-dark matter ratio
        (right) contour plots for a slice of 0.26 h$^{-1}$
        Mpc thickness at $z=0$. The solid contours represent,
        respectively, regions with $T>10^5$ K (left) and
        $\rho_{b}/\rho > f_c$ (right). The dotted lines represent
        $T<10^5$ K (left) and $\rho_{b}/\rho< f_c$ (right).}

\end{figure*}

\newpage
\begin{figure*}[hbt]
\centerline{\psfig{figure=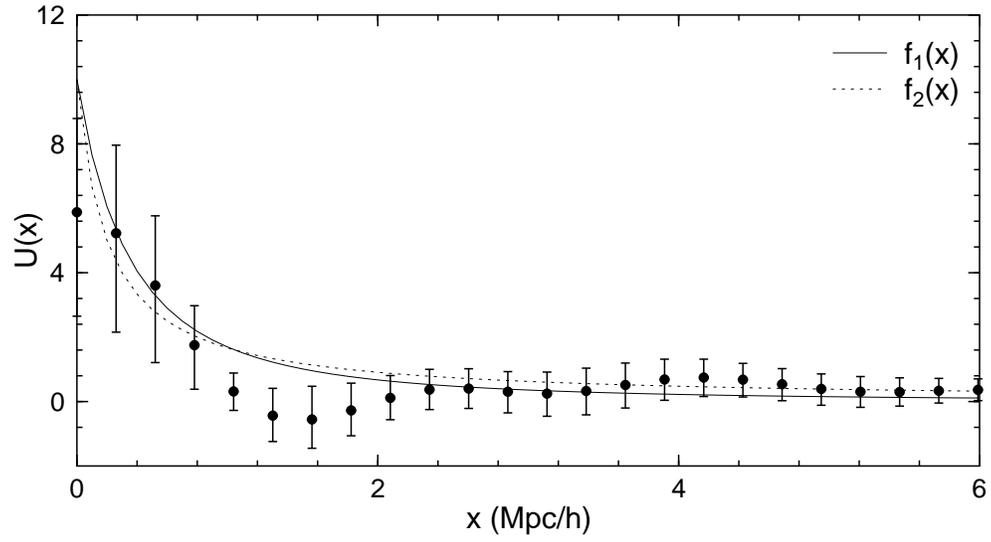}}

\caption{Mass density profile $U(x)$ in physical space.
         The fitting curves are given by equation (17) with
         $(\alpha,\beta)= (1,3)$ (solid line) and $(1,2)$ (dashed line).}

\end{figure*}

\end{document}